\documentclass[conference,letterpaper]{IEEEtran}

\usepackage[margin=0.55in]{geometry}
\usepackage[utf8]{inputenc} 
\usepackage[T1]{fontenc}
\usepackage{ifthen}
\usepackage{cite}
\usepackage[cmex10]{amsmath} 
\usepackage{amsthm,amsfonts,amssymb}
\usepackage{graphicx}
\usepackage{subfig}
\usepackage{blkarray}
\usepackage[mathscr]{euscript}
\usepackage{enumitem}
\usepackage{blkarray}
\usepackage{stfloats}%
\newtheorem{theorem}{Theorem}[section]
\newtheorem{corollary}{Corollary}[section]
\newtheorem{proposition}{Proposition}[section]
\newtheorem{lemma}[section]{Lemma}
\theoremstyle{remark}
\newtheorem*{remark}{Remark}
\theoremstyle{definition}
\newtheorem{definition}{Definition}[section]
\usepackage{url}
\usepackage{arydshln}
\usepackage{mathtools}
\usepackage{dsfont}

\begin{document}
\title{Bounds on the Feedback Capacity of the $(d,\infty)$-RLL Input-Constrained Binary Erasure Channel} 


\author{%
  \IEEEauthorblockN{V.~Arvind Rameshwar}
 \IEEEauthorblockA{Department of Electrical Communication Engineering\\
 	Indian Institute of Science, Bengaluru\\
 	Email: \texttt{vrameshwar@iisc.ac.in}}
 \and
 \IEEEauthorblockN{Navin Kashyap}
 \IEEEauthorblockA{Department of Electrical Communication Engineering\\
 	Indian Institute of Science, Bengaluru\\
 	Email: \texttt{nkashyap@iisc.ac.in}
 }
\thanks{The work of N.~Kashyap was supported in part by MATRICS grant \ MTR/2017/000368 from the Science and Engineering Research Board (SERB), Govt. of India. The work of V.~A.~Rameshwar was supported by a Qualcomm Innovation Fellowship, India.  The authors would like to thank O.~Sabag for having shared the code for the numerical computation of the upper bounds.}
}

\IEEEoverridecommandlockouts


\maketitle

\begin{abstract}
   The paper considers the input-constrained binary erasure channel (BEC) with causal, noiseless feedback. The channel input sequence respects the $(d,\infty)$-runlength limited (RLL) constraint, i.e., any pair of successive $1$s must be separated by at least $d$ $0$s. We derive upper and lower bounds on the feedback capacity of this channel, for all $d\geq 1$, given by:\\ $\max\limits_{\delta \in [0,\frac{1}{d+1}]}R(\delta) \leq C^{\text{fb}}_{(d\infty)}(\epsilon) \leq \max\limits_{\delta \in [0,\frac{1}{1+d\epsilon}]}R(\delta)$, where the function $R(\delta) = \frac{h_b(\delta)}{d\delta + \frac{1}{1-\epsilon}}$, with $\epsilon\in [0,1]$ denoting the channel erasure probability, and $h_b(\cdot)$ being the binary entropy function. We note that our bounds are tight for the case when $d=1$ (see Sabag et al. (2016)), and, in addition, we demonstrate that for the case when $d=2$, the feedback capacity is equal to the capacity with non-causal knowledge of erasures, for $\epsilon \in [0,1-\frac{1}{2\log(3/2)}]$. For $d>1$, our bounds differ from the non-causal capacities (which serve as upper bounds on the feedback capacity) derived in Peled et al. (2019) in only the domains of maximization. The approach in this paper follows Sabag et al. (2017), by deriving single-letter bounds on the feedback capacity, based on output distributions supported on a finite $Q$-graph, which is a directed graph with edges labelled by output symbols. 
\end{abstract}


\section{Introduction}

Memoryless channels were introduced by Shannon \cite{Sh48} as models for communication links, and have, since then, been the object of much research activity in information theory. The capacity of a memoryless channel has an elegant, single-letter expression, $C = \sup_{P(x)}I(X;Y)$, and this expression is computable for a wide range of channels (see \cite{Blahut}, \cite{Arimoto}). Furthermore, it was shown by Shannon \cite{Sh56} that causal, noiseless feedback of the outputs does not increase the capacity of memoryless channels.

\begin{figure}[!t]
	\centering
	\includegraphics[width=0.48\textwidth]{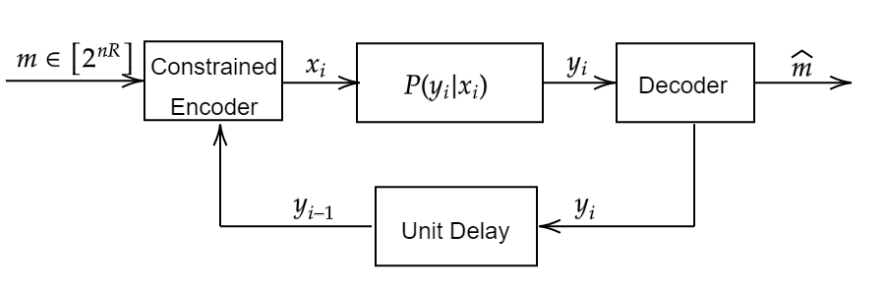}
	\caption{System model of an input-constrained DMC with causal, noiseless feedback.}
	\label{fig:genconstdmc}
\end{figure}

The case of discrete channels with memory (or discrete finite-state channels or FSCs) is quite different. For the purposes of this discussion, we restrict our attention to discrete memoryless channels with input constraints, which are specific instances of FSCs, and which find application in magnetic and optical recording systems \cite{Roth}, \cite{Immink}. Figure \ref{fig:genconstdmc} shows a generic discrete memoryless channel input constraints, with causal, noiseless feedback present. In this work, we shall focus on the binary erasure channel (or BEC) (shown in Figure \ref{fig:bec}), the inputs of which are constrained to obey the $(d,\infty)$-runlength limited (RLL) input constraint\footnote{The $(d,\infty)$-RLL constraint is a special case of the $(d,k)$-RLL constraint on binary sequences, which requires that the length of any run of consecutive $0$s is at least $d$ and at most $k$.}, which mandates that there needs to be at least $d$ $0$s between any pair of successive $1$s. As was shown in the works \cite{SabBEC}, \cite{Hanerasure}, and \cite{SabBIBO}, the capacities of the $(1,\infty)$-RLL input-constrained binary erasure and binary symmetric channels are strictly larger than their respective non-feedback capacities, for at least some values of the channel parameters. Hence, Shannon's argument does not apply to DMCs with constrained inputs, and special tools are required to determine the feedback capacities of such channels.

In the context of the BEC with input constraints, the feedback capacities of the $(1,\infty)$- and the $(0,k)$-RLL input constraints (for $k\geq 1$) were computed in the works \cite{SabBEC} and \cite{0k}, respectively. Both of these works use the capacity of the BEC with non-causal knowledge of erasures as an upper bound on the feedback capacity, and put down an explicit coding scheme, whose rate is a lower bound on the feedback capacity, and demonstrate that the upper and lower bounds match. Interestingly, this gives rise to the fact that the non-causal capacities of the BEC with the $(1,\infty)$- and the $(0,k)$-RLL input constraints are equal to the respective feedback capacities. Moreover, it was shown in \cite{0k} that the feedback capacity of the $(2,\infty)$-RLL input-constrained BEC is less than the corresponding non-causal capacity, when the channel parameter, $\epsilon$, is equal to $0.5$. The proof involved deriving an upper bound on the feedback capacity, and proving that the upper bound is strictly less than the non-causal capacity at $\epsilon=0.5$. In this paper, we show that for the $(2,\infty)$-RLL input-constrained BEC, the feedback capacity is in fact equal to the non-causal capacity for $\epsilon\in [0,1-\frac{1}{2\log(3/2)}]$. This result is a by-product of the main theorem in our paper, which presents an analytical lower bound on the feedback capacity of the $(d,\infty)$-RLL input-constrained BEC, for all $d$ greater than or equal to $1$. For the case when $d=1$, our lower bound is tight for all values of $\epsilon$ \cite{SabBEC}, and for $d=2$, our lower bound is numerically close to an upper bound derived using techniques in \cite{Graph_based}, for $\epsilon$ larger than $1-\frac{1}{2\log(3/2)}$, too. Our lower bounds are single-parameter optimization problems, and differ from the expression for the non-causal capacity (established for the $(d,\infty)$-RLL input-constrained BEC in \cite{0k}) in only the domain of maximization. 

Our techniques involve the use of the single-letter bounding techniques of Sabag et al. in \cite{Single}, and we come up with a specific $Q$-graph (or a $Q$-context mapping) and a choice of a specific ``BCJR-invariant'' input distribution, which gives rise to an achievable rate.

\section{Notation and Preliminaries}

\subsection{Notation}

In what follows, random variables will be denoted by capital letters, and their realizations by lower-case letters, e.g., $X$ and $x$, respectively. Calligraphic letters, e.g., $\mathscr{X}$, denote sets. We use the notation $[n]$ to denote the set, $\{1,2,\ldots,n\}$, of integers, and the notation $[a:b]$, for $a<b$, to denote the set of integers $\{a,a+1,\ldots,b\}$. The notations $X^{N}$ and $x^N$ denote the random vector $(X_1,\ldots,X_N)$ and the realization $(x_1,\ldots,x_N)$, respectively. We define $\mathbf{e}_i^n$ to be the vector $(e_0,e_1,\ldots,e_n)$, of length $n+1$, with $e_i=1$, and $e_j=0$, for $j\neq i$. Further, $P(x), P(y)$ and $P(y|x)$ are used to denote the probabilities $P_X(x), P_Y(y)$ and $P_{Y|X}(y|x)$, respectively. As is usual, the notations $H(X)$ and $I(X;Y)$ stand for the entropy of the random variable $X$, and the mutual information between the random variables $X$ and $Y$, respectively, and $h_b(p)$ is the binary entropy function, for $p\in [0,1]$. Finally, for a real number $\alpha \in [0,1]$, we define $\bar{\alpha}=1-\alpha$. All logarithms are taken to the base $2$.

\subsection{Problem Definition}

\begin{figure}[htbp]
	\centering
	\includegraphics[scale=0.6]{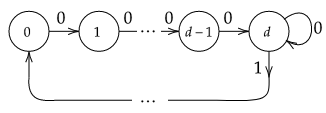}
	\caption{A deterministic presentation of the $(d,\infty)$-RLL input constraint. The nodes of the presentation represent the channel states and the labels on the edges represent inputs.}
	\label{fig:d_inf_constraint}
\end{figure}

The communication setting of an input-constrained memoryless channel with feedback is shown in Figure \ref{fig:genconstdmc}. A message $M$ is drawn uniformly from the set $\{1,2,\ldots,2^{nR}\}$, and is made available to the encoder. The encoder, at time $i$, also has access to noiseless feedback in the form of the outputs, $y^{i-1}$, from the decoder, and produces a binary input symbol $x_i \in \{0,1\}$, as a function of the specific instance of the message, $m$, and the previous outputs, $y^{i-1}$. The encoder is constrained in that the sequence of input symbols $x_1x_2x_3\ldots$ must satisfy the $(d,\infty)$-RLL input constraint, a deterministic presentation of which is shown in Figure \ref{fig:d_inf_constraint}. We set the channel state alphabet, $\mathscr{S}$, to be $\{0,1,\ldots,d\}$. Furthermore, the channel is memoryless in the sense that

\begin{equation*}
P(y_i|x^{i},y^{i-1}) = P(y_i|x_i), \quad \forall i.
\end{equation*}

\begin{figure}[htbp]
	\centering
	\includegraphics[scale=0.7]{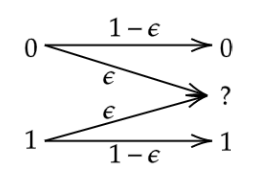}
	\caption{The binary erasure channel.}
	\label{fig:bec}
\end{figure}

Our focus is on the binary erasure channel, or the BEC, shown in Figure \ref{fig:bec}. Here, the input alphabet, $\mathscr{X} = \{0,1\}$, while the output alphabet is $\mathscr{Y} = \{0,?,1\}$, where $?$ denotes an erasure. Let $\epsilon \in [0,1]$ be the erasure probability of the channel.

\begin{figure*}[!b]
	\hrule
	\begin{align}
	B_{s^+}(\left(\gamma_q(s):s\in \mathscr{S}\right),y) = \frac{\sum_{x,s} \mathds{1}\{s^{+}=f(x,s)\}\gamma_q(s)P(x|s,q)P(y|x)}{\sum_{x^{\prime},s^{\prime}}\gamma_q(s^{\prime})P(x^{\prime}|s^{\prime},q)P(y|x^{\prime})}. \label{eq:bs}
	\end{align}
\end{figure*}

\begin{definition}
	An $(n,2^{nR},(d,\infty))$ feedback code for an input-constrained channel is defined by a set of encoding functions:
	\begin{equation*}
	f_i: \{1,\ldots, 2^{nR}\}\times \mathscr{Y}^{i-1} \rightarrow \mathscr{X}, \quad i\in [n],
	\end{equation*}
	which satisfy $f_i(m,y^{i-1}) = 0$, if $f_{(i-j)^+}(m, y^{(i-j-1)^+}) = 1$ (where $x^+$ is equal to   $\max\{x,0\}$), for some $j\in [d]$, and for all $(m,y^{i-1})$ , and a decoding function:
	\begin{equation*}
	\Psi: \mathscr{Y}^n \rightarrow \{1,\ldots,2^{nR}\}.
	\end{equation*}
	The average probability of error for a code is defined as $P_e^{(n)} = P(M\neq \Psi(Y^n))$. A rate $R$ is said to be $(d,\infty)$-achievable if there exists a sequence of $(n,2^{nR},(d,\infty))$ codes, such that $\lim_{n\rightarrow \infty} P_e^{(n)} = 0$. The capacity, $C_{(d,\infty)}^{\text{fb}}(\epsilon)$, is defined to be the supremum over all $(d,\infty)$-achievable rates, and is a function of the erasure probability, $\epsilon$.
\end{definition}

\subsection{$Q$-graphs and $(S,Q)$-graphs}
\label{sec:Q-graph}
We now recall the definitions of the $Q$-graph and the $(S,Q)$-graph introduced in \cite{Single}.
\begin{definition}
	A $Q$-graph is a finite irreducible labelled directed graph on a vertex set $\mathscr{Q}$, with the property that each $q\in \mathscr{Q}$ has at most $|\mathscr{Y}|$ outgoing edges, each labelled by a unique $y\in \mathscr{Y}$.
\end{definition}

\begin{figure}[htbp]
	\centering
	\includegraphics[width=0.37\textwidth]{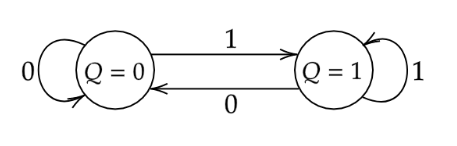}
	\caption{A $\mathscr{Q}$-graph where each node represents the last channel output, where $\mathscr{Y}=\{0,1\}$. The labels on the edges represent outputs.}
	\label{figQ}
\end{figure}

Thus, there exists a function $\Phi: \mathscr{Q} \times \mathscr{Y} \rightarrow \mathscr{Q}$, such that $\Phi(q,y)=q'$ iff there is an edge $q \stackrel{y}{\longrightarrow} q'$ in the $\mathscr{Q}$-graph.
We arbitrarily label one vertex of the $\mathscr{Q}$-graph as $q_0$. For any positive integer $n$, there is a one-to-one correspondence between sequences in $(y_1,y_2,\ldots,y_n) \in \mathscr{Y}^n$ and directed paths in the $\mathscr{Q}$-graph  starting from $q_0$: $q_0 \stackrel{y_1}{\longrightarrow} q_1 \stackrel{y_2}{\longrightarrow}
\cdots \stackrel{y_n}{\longrightarrow} q_n$. Figure \ref{figQ} depicts an example of a $\mathscr{Q}$-graph.

\begin{definition}
	Given an input-constrained DMC specified by $\{P(y|x)\}$ and the states of the presentation of the input constraint of which obey $s_t=f(s_{t-1},x_t)$, and a $\mathscr{Q}$-graph with vertex set $\mathscr{Q}$, the $(\mathscr{S},\mathscr{Q})$-graph is defined to be a directed graph on the vertex set $\mathscr{S}\times \mathscr{Q}$, with edges $(s,q) \xrightarrow{(x,y)} (s^{\prime},q^{\prime})$ if and only if $P(y|x)>0$, $s^{\prime}=f(s,x)$, and $q^{\prime}=\Phi(q,y)$.
	\label{def:SQ_graph}
\end{definition}

Now, given an input distribution $\{P(x|s,q)\}$ defined for each $(s,q)$ in the $(\mathscr{S},\mathscr{Q})$-graph, we have a Markov chain on $\mathscr{S}\times \mathscr{Q}$, where the transition probability associated with any edge $(x,y)$ emanating from $(s,q)\in \mathscr{S}\times \mathscr{Q}$ is $P(y|x)P(x|s,q)$. Let $\mathscr{G}(\{P(x|s,q)\})$ be the subgraph remaining after discarding edges of zero probability. We then define
\begin{align*}
\Omega \triangleq \bigl\lbrace \{P(x | s,q)\}: \mathscr{G}& (\{P(x|s,q)\})\text{ has a single} \\ 
& \text{closed communicating class}\bigr\rbrace.
\end{align*}
Given an irreducible $\mathscr{Q}$-graph, an input distribution $\{P(x|s,q)\} \in \Omega$ is said to be \emph{aperiodic}, if the corresponding graph, $\mathscr{G}(\{P(x|s,q)\})$, is aperiodic. For such distributions, the Markov chain on $\mathscr{S}\times \mathscr{Q}$ has a unique stationary distribution $\pi(s,q)$.

An aperiodic input is said to be BCJR-invariant, if its state probability vector, $\left(\pi(S=s|Q=q): s\in \mathscr{S}\right)$, satisfies
\begin{equation*}
\pi(S=s^{+}|Q=\Phi(q,y)) = B_{s^+}(\left(\pi(s|q):s\in \mathscr{S}\right),y),
\end{equation*}
for all $s,q \in \mathscr{S}\times \mathscr{Q}$, and $y\in \mathscr{Y}$, where the function \\$B_s: \Delta(|\mathscr{S}|-1)\times \mathscr{Y}\rightarrow [0,1]$ is shown in equation \eqref{eq:bs}, with $\Delta(|\mathscr{S}|-1)$ denoting the $(|\mathscr{S}|-1)$-dimensional unit simplex, i.e., $\Delta(|\mathscr{S}|-1) = \{\mathbf{u}\in [0,1]^{|\mathscr{S}|}: \sum_{i=1}^{|\mathscr{S}|}u_i = 1\}$. Note that $\left(\gamma_q(s):s\in \mathscr{S}\right)$ is an arbitrary element of 
$\Delta(|\mathscr{S}|-1)$ that is associated with node $q$ of the $Q$-graph.

\subsection{Bounds on Feedback Capacity}
We shall make use of the following single-letter bounds on feedback capacity (specialized to input-constrained DMCs) of \cite{Single}. The theorem below provides a lower bound on $C^{\text{fb}}_{DMC}$ by considering input distributions that are BCJR-invariant.

\begin{theorem}[\cite{Single}, Theorem 3]
	\label{thm:LB}
	The feedback capacity, $C^{\text{fb}}_{DMC}$, of input-constrained DMCs is lower bounded as
	\begin{equation*}
	C^{\text{fb}}_{DMC} \geq I(X;Y|Q),
	\end{equation*}
	for all aperiodic input distributions, $\{P(x|s,q)\} \in \Omega$, which are BCJR-invariant. The random variables $X, Y, S, Q$ are associated with the time invariant system, with their joint distribution given by $P(x,s,y,q) = \pi(s,q)P(x|s,q)P(y|x)$, where $\pi(s,q)$ is the stationary distribution of the $(S,Q)$-graph.
\end{theorem}

The next theorem, which is a specialization of Theorem 2 of \cite{Single}, provides a single-letter upper bound on $C^{\text{fb}}_{\text{DMC}}$, when the presentation of the input constraint is assumed to be irreducible:

\begin{theorem}[\cite{Single}, Theorem 2]\label{thm:UB}
	The feedback capacity, $C^{\text{fb}}_{DMC}$, of input-constrained DMCs, when the presentation of the input-constraint is taken to be irreducible, is upper bounded as
	\begin{equation*}
	C^{\text{fb}}_{DMC} \leq \sup_{P(x|s,q)\in \Omega} I(X;Y|Q),
	\end{equation*}
	for all irreducible $Q$-graphs with $q_0$ such that $(s_0,q_0)$ lies in an aperiodic closed communicating class.
\end{theorem}

\begin{remark}
	We implictly assume that the encoder and the decoder know the initial channel state, $s_0$.
\end{remark}

\section{Main Results}
The following theorem states our main result concerning the capacity of the $(d,\infty)$-RLL input-constrained BEC with feedback. We shall make use of the function, $R(\delta)$, defined for $\delta\in [0,1]$, which is given by
\[
R(\delta) = \frac{h_b(\delta)}{d\delta + \frac{1}{1-\epsilon}}.
\]

\begin{theorem} \label{thm:main}
	The feedback capacity of the $(d,\infty)$-RLL input-constrained BEC, for $\epsilon>0$, is lower bounded as follows:
	\begin{equation*}
	C_{(d,\infty)}^{\text{fb}}(\epsilon) \geq \max\limits_{\delta \in [0,\frac{1}{d+1}]} R(\delta).
	\end{equation*}
	
\end{theorem}
\begin{remark}
	At $\epsilon=0$, the capacities with and without feedback are identical, and are given by $C(0) = \max\limits_{\delta \in [0,1]} \frac{h_b(\delta)}{d\delta + 1}$.
\end{remark}
\begin{remark}
	The capacity of the $(d,\infty)$-RLL input-constrained BEC with non-causal knowledge of erasures, which serves as an upper bound on the feedback capacity, is given by \cite[Lemma 7]{0k}:
	\begin{equation*}
	C_{(d,\infty)}^{\text{nc}}(\epsilon) = \max\limits_{\delta \in [0,\frac{1}{2}]} R(\delta).
	\end{equation*}
	By standard calculus arguments, it holds that $R(\cdot)$ is concave, and, hence, has a unique maximum in $[0,1]$. Indeed, since $R^{\prime}(0^+)>0$ and $R^{\prime}\left(\frac{1}{2}\right)<0$, the unique maximum must lie in the interval $[0,\frac{1}{2}]$, where $R^\prime(\cdot)$ denotes the derivative of $R(\cdot)$. Note that the difference between the lower bound in Theorem \ref{thm:main} and the non-causal capacity expression is only in the domain of maximization.
	
\end{remark}
Theorem \ref{thm:main} follows from the construction of a family of $Q$-graphs, indexed by $d$, and the identification of a  BCJR-invariant input distribution. The construction of the $Q$-graphs and the input distribution is the subject of Section \ref{sec:Q_const}. Theorem \ref{thm:main}, whose proof is provided in Section \ref{sec:proof_thm}, then follows from Theorem \ref{thm:LB} by evaluating the conditional mutual information, $I(X;Y|Q)$, using the BCJR-invariant distribution identified.

Theorem \ref{thm:main} implies the following corollaries:

\begin{corollary}\label{coroll}
	
	\begin{enumerate}
		\item For $\epsilon \leq 1-\frac{1}{2\log(\frac{3}{2})}$, the feedback capacity of the $(2,\infty)$-RLL input-constrained BEC is equal to the non-causal capacity, and is given by:
		\begin{equation*}
		C_{(2,\infty)}^{\text{fb}}(\epsilon) = C_{(2,\infty)}^{\text{nc}}(\epsilon) =  \max\limits_{\delta \in [0,\frac{1}{2}]} R(\delta).
		\end{equation*}
		Further, for $\epsilon > 1-\frac{1}{2\log(\frac{3}{2})}$, the feedback capacity is lower bounded as:
		\begin{equation*}
		C_{(2,\infty)}^{\text{fb}}(\epsilon) \geq R\left(\frac{1}{3}\right).
		\end{equation*}
		\item For $d\geq 3$, and for $\epsilon>0$, the feedback capacity of the $(d,\infty)$-RLL input-constrained BEC is lower bounded as:
		\begin{equation*}
		C_{(d,\infty)}^{\text{fb}}(\epsilon) \geq R\left(\frac{1}{d+1}\right).
		\end{equation*}
		Moreover, the non-causal capacity is strictly larger than the lower bound, in this case.
	\end{enumerate}
\end{corollary}
The proofs of the corollaries are relegated to Appendix B.

\begin{figure*}[t]
	\centering
	\includegraphics[width=0.7\textwidth]{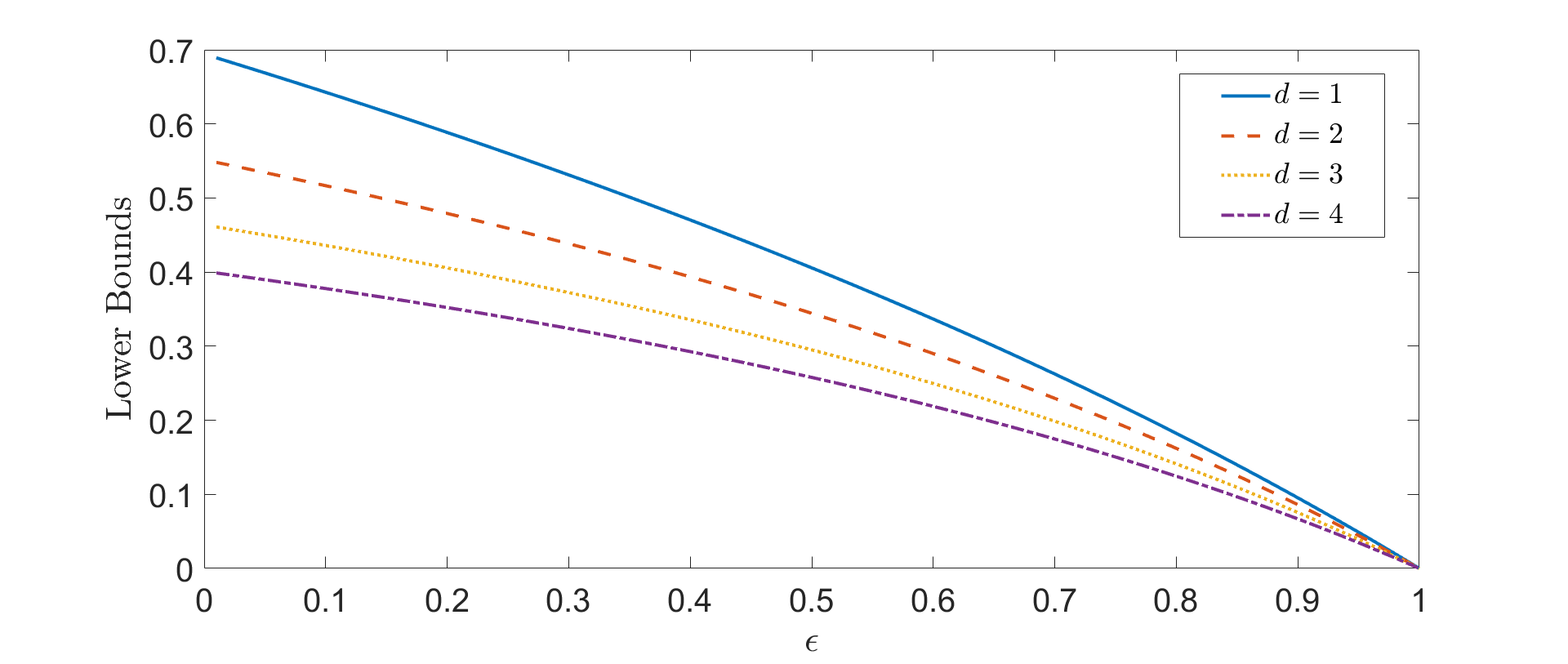}
	\caption{Plots of our lower bounds for $d=1,2,3,4$.}
	\label{fig:plotdinf}
\end{figure*}

Figure \ref{fig:plotdinf} shows a plot of our lower bounds, for different values of $d$. We note that for $d=1$, our lower bound is tight \cite{SabBEC}. However, for $d\geq 3$, our lower bounds are not tight in the neighbourhood of $\epsilon=0$. 

The following result provides an upper bound on the feedback capacity of the $(d,\infty)$-RLL input-constrained BEC, by considering a specific family of $Q$-graphs, each graph in which has $d+1$ nodes, and invoking Theorem \ref{thm:UB}.

\begin{proposition}\label{lemma:UB}
	The feedback capacity of the $(d,\infty)$-RLL input-constrained BEC is upper bounded as:
	\begin{equation}
	\label{eq:UB}
	C_{(d,\infty)}^{\text{fb}}(\epsilon) \leq \max\limits_{\delta \in [0,\frac{1}{1+d\epsilon}]} R(\delta).
	\end{equation}
	
	Further, for values of $\epsilon$ larger than $\epsilon^{\star}$, where $\epsilon^{\star}$ satisfies
	\[
	(d\epsilon)^{\left(\frac{1}{1-\epsilon}+d\right)} = (1+d\epsilon)^d,
	\]
	the upper bound in \eqref{eq:UB} is less than the non-causal capacity.
\end{proposition}

\begin{figure*}[t]
	\centering
	\includegraphics[width=0.7\textwidth]{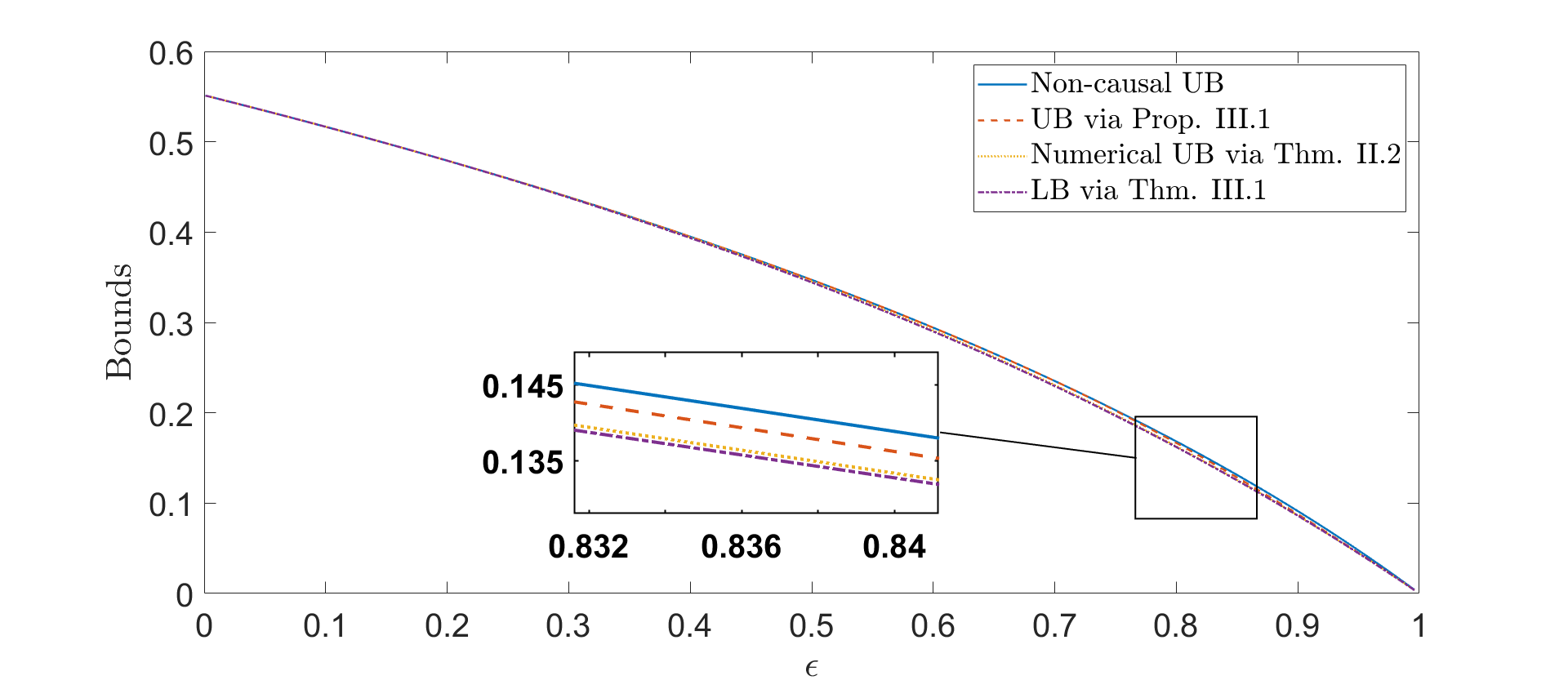}
	\caption{Plot comparing bounds for the $(2,\infty)$-RLL input-constrained BEC.}
	\label{fig:UB_2_inf}
\end{figure*}

\begin{figure*}[t]
	\centering
	\includegraphics[width=0.7\textwidth]{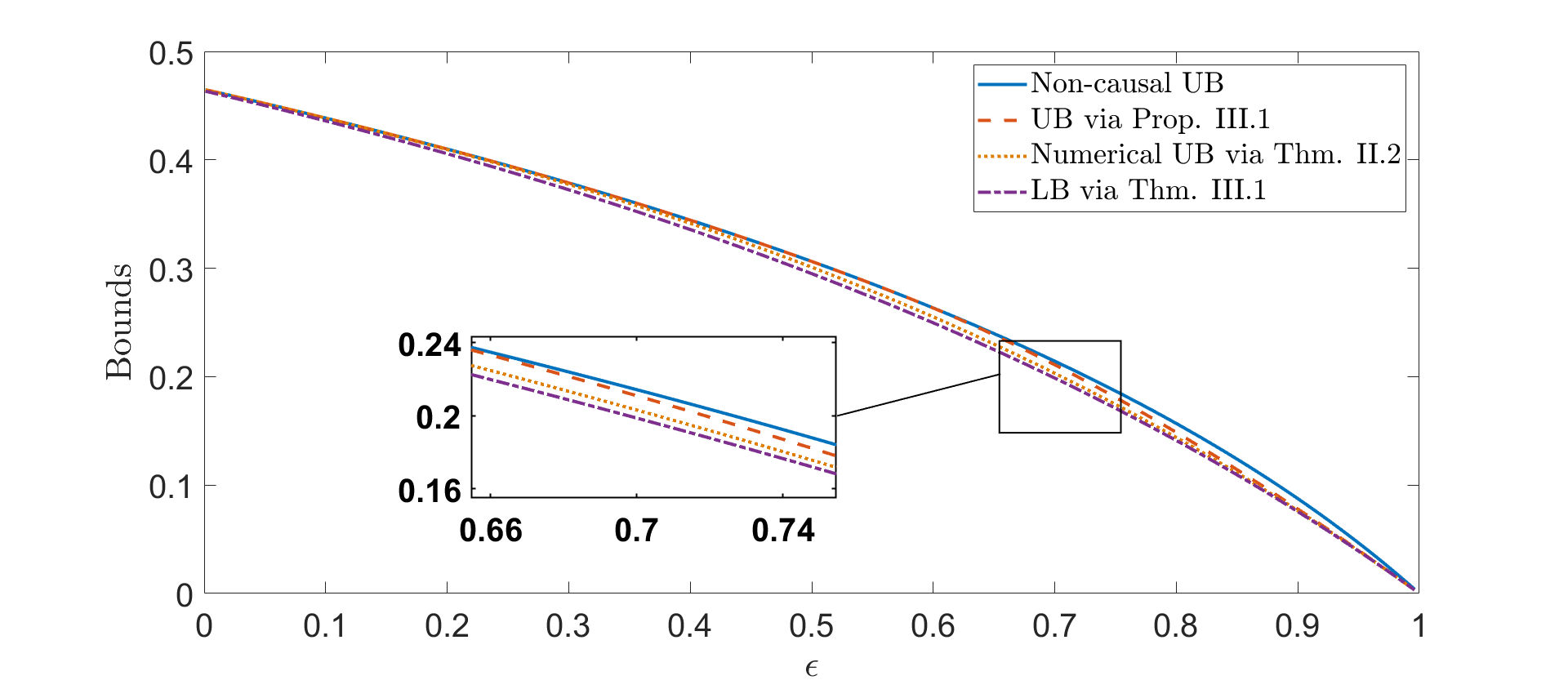}
	\caption{Plot comparing bounds for the $(3,\infty)$-RLL input-constrained BEC.}
	\label{fig:UB_3_inf}
\end{figure*}

The proof of Proposition \ref{lemma:UB} is provided in Section \ref{sec:proof_UB}. Figures \ref{fig:UB_2_inf} and \ref{fig:UB_3_inf} show, for $d=2$ and $d=3$, respectively, comparisons of the bounds in Theorem \ref{thm:main} and Proposition \ref{lemma:UB} with the non-causal capacity, and numerical evaluations of an upper bound obtained using Theorem \ref{thm:UB}, by employing the same $Q$-graphs that were used for our lower bounds. The numerical upper bounds were computed by casting the upper bound in Theorem \ref{thm:UB} as a convex programming problem, as was discussed in \cite{Graph_based}. It is clear from the plots that the upper bound in Proposition \ref{lemma:UB} is tighter than the non-causal capacity upper bound, for values of $\epsilon$ larger than $\epsilon^{\star}$---for example, for $d=2$, $\epsilon^{\star}\approx 0.6960$, and for $d=3$, $\epsilon^{\star}\approx 0.5850$. Futher, for $d=2$, we note that the upper and lower bounds are equal for $\epsilon\in [0,1-\frac{1}{2\log(3/2)}]$, and are numerically close in value, for $\epsilon> 1-\frac{1}{2\log(3/2)}$.


%

\section{Construction of Q-graph and Input Distribution}
\label{sec:Q_const}
We shall henceforth consider $\mathscr{X} = \{0,1\}, \mathscr{Y} = \{0,?,1\}$, and $\mathscr{S} = \{0,1,\ldots,d\}$ to be the input, output and state alphabets, respectively, for the $(d,\infty)$-RLL input constrained BEC.

In this section, we shall construct a family of $Q$-graphs, indexed by the value of $d$, and identify a BCJR-invariant input distribution on the $(S,Q)$-graph corresponding to each $Q$-graph in the family.

\subsection{Construction of Q-graphs}
%
%
%
With each value of $d$ corresponding to the $(d,\infty)$-RLL input-constrained BEC, we associate a $Q$-graph, \\$\mathscr{G}_{Q}^{(d)} = (\mathscr{V}^{(d)},\mathscr{E}^{(d)},\mathscr{L}^{(d)})$, with vertex set $\mathscr{V}^{(d)}$, edge multi-set $\mathscr{E}^{(d)}$, and a labelling function $\mathscr{L}^{(d)}: \mathscr{E}^{(d)}\rightarrow \{0,?,1\}$.

We write $\mathscr{V}^{(d)} = \mathscr{V}_{\text{DB}}^{(d)} \cup \{Q_0, Q_1,\ldots,Q_{d-1}\}$, where $\mathscr{V}_{\text{DB}}^{(d)}$ is constructed as follows. Each vertex in $\mathscr{V}_{\text{DB}}^{(d)}$ is uniquely identified by a $d$-tuple over $\{0,?\}$, i.e., each vertex in $\mathscr{V}_{\text{DB}}^{(d)}$ can alternatively be written as $\mathbf{q} = (w_{0}^q,w_{1}^q,\ldots,w_{d-1}^q)$, where $w^q_i \in \{0,?\}$, for $i\in \{0,1,\ldots,d-1\}$. In what follows, we shall suppress the superscript $q$, as it will be clear from the context. Note that there are $2^d$ vertices in $\mathscr{V}_{\text{DB}}^{(d)}$. Let us denote by $Q_d$, the vertex $(0,0,\ldots,0)\in \mathscr{V}_{\text{DB}}^{(d)}$.

We shall now construct the edge multi-set, $\mathscr{E}^{(d)}$. Firstly, for vertices in $\mathscr{V}_{\text{DB}}^{(d)}$, we define the set 
\begin{align*}
	\mathscr{E}_{\text{DB}}^{(d)} = \lbrace\mathbf{e}:\mathbf{e} = &\left((w_{0},w_{1}\ldots,w_{d-1}),(w,w_{0},w_1,\ldots,w_{d-2})\right),\\
	& \text{ for }(w_{0},\ldots,w_{d-1}), \ (w,w_{0},\ldots,w_{d-2}) \in \mathscr{V}_{\text{DB}}^{(d)} \rbrace.
\end{align*}

Further, for any edge \\$\mathbf{e} = \left((w_{0},w_{1}\ldots,w_{d-1}),(w,w_{0},w_{2},\ldots,w_{d-2})\right)$, we define the labelling function $\mathscr{L}_{\text{DB}}^{(d)}: \mathscr{E}_{\text{DB}}^{(d)}\rightarrow \{0,?\}$ such that $\mathscr{L}_{\text{DB}}^{(d)}(\mathbf{e}) = w$.



The edge set, $\mathscr{E}^{(d)}$, is constructed by adding to the set $\mathscr{E}_{\text{DB}}^{(d)}$, the edges $(\mathbf{q},Q_0)$, for $\mathbf{q}\in \mathscr{V}_{\text{DB}}^{(d)}$, and, corresponding to each $Q_i$, $i\in [0:d-1]$, two edges, $\mathbf{e}^{Q_i}_1$ and $\mathbf{e}^{Q_i}_2$, from $Q_i$ to $Q_{i+1}$. Note that the $Q$-graph, $\mathscr{G}_Q^{(d)}$, has parallel edges, which are precisely the edges from $Q_i$ to $Q_{i+1}$, for $i\in [0:d-1]$.


Finally, the labelling function, $\mathscr{L}^{(d)}:\mathscr{E}^{(d)}\rightarrow \{0,?,1\}$ is defined as:
\begin{equation}\label{eq:label}
\mathscr{L}^{(d)}(\mathbf{e})=
\begin{cases*}
\mathscr{L}^{(d)}_{\text{DB}}(\mathbf{e}),\text{ if $\mathbf{e}\in \mathscr{E}^{(d)}_{\text{DB}}$},\\
1,\text{ if $\mathbf{e}=(\mathbf{q},Q_0)$ for $\mathbf{q}\in \mathscr{V}^{(d)}_{\text{DB}}$},\\
0, \text{ if $\mathbf{e} = \mathbf{e}^{Q_i}_1$ for some $i\in [0:d-1]$},\\
?, 	\text{ if $\mathbf{e} = \mathbf{e}^{Q_i}_2$ for some $i\in [0:d-1]$}.
\end{cases*}
\end{equation}

\begin{remark}
	The subscript ``DB'' refers to the fact that the subgraph $\mathscr{G}_{DB}^{(d)} = (\mathscr{V}_{\text{DB}}^{(d)},\mathscr{E}_{\text{DB}}^{(d)},\mathscr{L}_{\text{DB}}^{(d)})$, of $\mathscr{G}_{Q}^{(d)}$, is a de Bruijn graph of order $d$, on $\{0,?\}$ symbols.
\end{remark}

\begin{figure*}[t]
	\centering
	\subfloat[]{\includegraphics[scale=0.6]{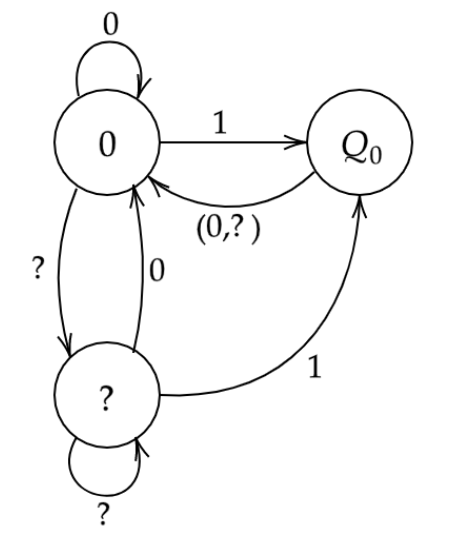}\label{fig:Q_1}}
	\hfill
	\subfloat[]{\includegraphics[scale=0.85]{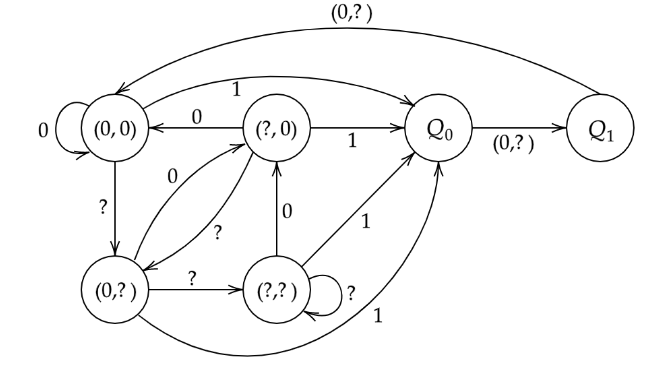}\label{fig:Q_2}}
	\caption{The $Q$-graphs $\mathscr{G}^{(1)}_Q$ and $\mathscr{G}^{(2)}_Q$ are shown above in (a) and (b), respectively. Note that the nodes in the de Bruijn components are labelled by $d$-tuples, with $Q_1 = 0$, in $\mathscr{G}^{(1)}_Q$, and $Q_2 = (0,0)$, in $\mathscr{G}^{(2)}_Q$. The parallel edges defined are here represented as a single edge with the label $(0,?)$, according to \eqref{eq:label}.}
\end{figure*}

We note that $\mathscr{G}_Q^{(d)}$ satsifies the properties of a $Q$-graph detailed out in Section \ref{sec:Q-graph}. The $Q$-graphs, $\mathscr{G}^{(1)}_Q$ and $\mathscr{G}^{(2)}_Q$ are shown in Figures \ref{fig:Q_1} and \ref{fig:Q_2}, respectively.

Our family of $Q$-graphs is then given by $\mathscr{H}:=\{\mathscr{G}^{(d)}_Q: d\in \mathbb{N}\}$, with the $Q$-graph $\mathscr{G}^{({d})}_Q$ being used to derive a lower bound for the $({d},\infty)$-RLL input-constrained BEC.
	
\begin{remark}
	The nodes $Q_0, Q_1,\ldots, Q_d$ of $\mathscr{G}^{(d)}_Q$ can be interpreted as follows. Let us define, for a fixed output sequence, $y^t$, the length-$(d+1)$ belief vector $\mathbf{z}_t:=\left(P(S_t=s_t|y^t):s_t\in \mathscr{S}\right)$. Note that the vector,  $\mathbf{z}_t$, which is an element of $\Delta(d)$, is precisely the state vector of the dynamic programming (DP) formulation of the feedback capacity problem \cite{Weiss}. The node $Q_i$ corresponds to the DP state $\mathbf{z}_t = \mathbf{e}_{i}^{d}$, for $i\in [0:d]$. This, in turn, has the interpretation that the nodes $Q_0,\ldots, Q_d$ represent the belief vectors when the decoder knows the channel state exactly.
	
\end{remark}

\subsection{Construction of a BCJR-invariant input distribution}
\label{subsec:BCJR_invar}

Given the $Q$-graph, $\mathscr{G}_Q^{(d)}$, for the $(d,\infty)$-RLL input constrained BEC, we shall call its corresponding $(S,Q)$ graph, $\mathscr{G}_{SQ}^{(d)}$, which is constructed following the description in definition \ref{def:SQ_graph}, with the presentation of channel states given in figure \ref{fig:d_inf_constraint}.

%
%

In this subsection, we shall construct a BCJR-invariant input distribution, $\left(\left(\tilde{P}(x|s,v):x\in \mathscr{X}\right): (s,v)\in \mathscr{S}\times \mathscr{V}^{(d)}\right)$. In tandem, we will also identify the conditional distribution, $\left(\tilde{\pi}(s|v):s\in \mathscr{S}\right)$, for each $v\in \mathscr{V}^{(d)}$, where $\tilde{\pi}(s|v) = \frac{\tilde{\pi}(s,v)}{\tilde{\pi}(v)}$, with $\left(\tilde{\pi}(s,v):s\in \mathscr{S},\ v\in \mathscr{V}^{(d)}\right)$ denoting the stationary distribution on the nodes of $\mathscr{G}_{SQ}^{(d)}$. Note that $\tilde{\pi}(v) = \sum_{s\in \mathscr{S}}\tilde{\pi}(s,v)$.




Since our input distribution, $\left(\tilde{P}(\cdot|s,v):s\in \mathscr{S},\ v\in \mathscr{V}^{(d)}\right)$, must respect the $(d,\infty)$ input constraint, we need that $\tilde{P}(1|s,v)=0$, for $s\in \{0,1,\ldots,d-1\}$, and we hence need only specify $\tilde{P}(1|d,v)$, for each $v\in \mathscr{V}^{(d)}$, to completely determine the input distribution.

To this end, we first associate with every node $v\in \mathscr{V}^{(d)}$, a probability vector, $\boldsymbol{\theta}_v:=\left(\theta_v(s):s\in \mathscr{S}\right)$, of length $d+1$. As we shall see, the vector $\boldsymbol{\theta}_v$ is a proxy for $\left(\tilde{\pi}(s|v):s\in \mathscr{S}\right)$ for a suitably defined input distribution.

We set $\boldsymbol{\theta}_{Q_i} = \mathbf{e}_{i}^{d}$, for $i\in [0:d]$. Further, in compliance with the $(d,\infty)$-RLL input constraint, we set $a_{Q_i}:=\tilde{P}(1|d,Q_i) = 0$, for $i\neq d$. Let us pick $a_{Q_d} := \tilde{P}(1|d,Q_d)$, such that $a_{Q_d}\leq \frac{1}{d+1}$. The reason for the restriction of $a_{Q_d}$ to the interval $[0,\frac{1}{d+1}]$ will be made clear in Lemma \ref{lemma:domain}. We shall, for conciseness of notation, represent $a_{Q_d}$ as $a$.

Now, for $\mathbf{q} = (w_{0},w_{1},\ldots,w_{d-1})\in \mathscr{V}^{(d)}_{\text{DB}}$, we define an accompanying vector $\boldsymbol{\beta}=(\beta_{0},\beta_{1},\ldots,\beta_{d-1})$, with\\ $\beta_s := |\{i\in [0:s-1]:w_i=0\}|$. In words, $\beta_s$ is the number of $0$s that appear strictly to the left of position $s$ in the $d$-tuple corresponding to $\mathbf{q}$. We then define $\boldsymbol{\theta}_{\mathbf{q}}$ as follows:

For $0\leq s\leq d-1$,
\begin{equation}\label{eq:theta_even}
{\theta}_{\mathbf{q}}(s) = 
\begin{cases}
0, \ w_s=0,\\
\frac{a}{(1-a)^{\beta_s}}, \ w_s=\ ?,
\end{cases}
\end{equation}
and $\theta_{\mathbf{q}}(d) = 1-\sum_{s=0}^{d-1}\theta_{\mathbf{q}}(s)$. It is clear from our definition that $\boldsymbol{\theta}_{Q_d} = \mathbf{e}_{d}^{d}$, where $Q_d = (0,0,\ldots,0)\in \mathscr{V}_{\text{DB}}^{(d)}$.


Finally, for $\mathbf{q}\in \mathscr{V}_{\text{DB}}^{(d)}$, we let $a_{\mathbf{q}}:= \tilde{P}(1|d,\mathbf{q}) = \frac{a}{\theta_{\mathbf{q}}(d)}$, where $a$ was defined earlier. Note that in order for $a_{\mathbf{q}}$ to be a valid probability, it must lie in the interval $[0,1]$, a sufficient condition for which is provided by Lemma \ref{lemma:domain} below.

\begin{lemma}\label{lemma:domain}
	For $a\in [0,\frac{1}{d+1}]$, $a_{\mathbf{q}}\in [0,1]$, for $\mathbf{q}\in \mathscr{V}_{\text{DB}}^{(d)}$ .
\end{lemma}

The proof of this lemma is taken up in  Appendix A.

\begin{remark}
Note that this procedure determines the vector $\left(a_v:v\in \mathscr{V}^{(d)}\right)$, thereby specifying the entire input distribution $\left(\tilde{P}(\cdot|s,v):s\in \mathscr{S},\ v\in \mathscr{V}^{(d)}\right)$. Moreover, our input distribution is aperiodic if $a<1$ and $\epsilon\neq 1$, as then there exists a self-loop of positive probability, $\bar{a}\bar{\epsilon}$, from the node $(d,Q_d)$ to itself, in $\mathscr{G}_{SQ}^{(d)}$. From our choice of the interval, $[0,\frac{1}{d+1}]$, in which $a$ lies, we see that $a$ indeed satisfies the condition that $a<1$, and we assume, further, that $\epsilon\neq 1$, as the feedback capacity at $\epsilon=1$ is well-known to be $C^{\text{fb}}(1) = 0$.
\end{remark}

Before we set out to prove that the aperiodic input distribution $\tilde{P}$ is indeed BCJR-invariant, we note that for all $s\in \mathscr{S}$ and $v\in \mathscr{V}^{(d)}$, the value $B_s(\boldsymbol{\theta}_v,y)$ (where the function $B_s: \Delta(d)\times \mathscr{Y} \rightarrow \Delta(d)$ was defined in equation \eqref{eq:bs}) can be computed explicitly for the $(d,\infty)$-RLL input constrained BEC with the input distribution $\tilde{P}$. 

For $0\leq s\leq d-1$, and for $\mathbf{q}\in \mathscr{V}_{\text{DB}}^{(d)}$, the following set of equations holds true:
\begin{equation}\label{eq:Bs0}
B_s(\boldsymbol{\theta}_\mathbf{q},0) = 
\begin{cases}
0, \ s=0,\\
\frac{\theta_{\mathbf{q}}(s-1)}{1-a_\mathbf{q}\theta_{\mathbf{q}}(d)} = \frac{\theta_{\mathbf{q}}(0)}{1-a}, \ 1\leq s \leq d-1,
\end{cases}
\end{equation}

\begin{equation}\label{eq:Bs?}
B_s(\boldsymbol{\theta}_\mathbf{q},?) = 
\begin{cases}
a_\mathbf{q}\theta_{\mathbf{q}}(d) = a, \ s=0,\\
\theta_{\mathbf{q}}(s-1), \ 1\leq s \leq d-1,
\end{cases}
\end{equation}

\begin{equation}\label{eq:Bs1}
B_s(\boldsymbol{\theta}_\mathbf{q},1) = 
\begin{cases}
1, \ s=0,\\
0, \ 1\leq s \leq d-1,
\end{cases}
\end{equation}
with $B_d(\boldsymbol{\theta}_\mathbf{q},y) = 1 - \sum_{s=0}^{d-1}B_s(\boldsymbol{\theta}_\mathbf{q},y)$, for $y\in \{0,?,1\}$. Note that the vector $\left(B_s(\boldsymbol{\theta}_\mathbf{q},1): s\in \mathscr{S}\right)$ is equal to $\mathbf{e}_0^{d}$.

Further, if $v=Q_i$, for some $i\in [0:d-1]$, it holds that
\begin{equation}\label{eq:BsQ}
\left(B_s(\boldsymbol{\theta}_v,y): s\in \mathscr{S}\right) = \mathbf{e}_{i+1}^{d},
\end{equation}
for $y\in \{0,?\}$.
We now note that the following lemma holds:

\begin{lemma}\label{lemma:bcjr}
	The set of probability vectors $\left(\boldsymbol{\theta}_{v}: v\in \mathscr{V}^{(d)}\right)$ along with the input distribution $\left(\tilde{P}(\cdot|s,v):s\in \mathscr{S},\ v\in \mathscr{V}^{(d)}\right)$ satisfies
	\begin{equation}\label{eq:bs_valid}
	\theta_{v^+}(s^+) = B_{s^+}(\boldsymbol{\theta}_v,y),
	\end{equation}
	for every $(s^+,v,y)\in \mathscr{S}\times \mathscr{V}^{(d)}\times \mathscr{Y}$, where $v^+ = \Phi(v,y)$, with $\Phi:\mathscr{V}^{(d)}\times \mathscr{Y}\rightarrow \mathscr{V}^{(d)}$ is the $Q$-node update function.
\end{lemma}
\begin{proof}
	Suppose first that $v \notin \mathscr{V}_{\text{DB}}^{(d)}$, and let $v=Q_i$, for some $i\in [0:d-1]$. Then, it follows from equation \eqref{eq:BsQ} that 
	\begin{align*}
	\mathbf{e}_{i+1}^{d}&=\left(B_s(\boldsymbol{\theta}_{Q_i},y):s\in \mathscr{S}\right) \\
	&= \boldsymbol{\theta}_{Q_{i+1}},
	\end{align*}
	for $y\in \{0,?\}$. Equation \eqref{eq:bs_valid} then follows by noting that $\Phi(Q_i,y) = Q_{i+1}$, for $y\in \{0,?\}$.
	
	Now, for $\mathbf{q}\in \mathscr{V}_{\text{DB}}^{(d)}$, when $y=1$, we see from equation \eqref{eq:Bs1} that
	\begin{align*}
	\mathbf{e}_{0}^{d}&=\left(B_s(\boldsymbol{\theta}_{v},1):s\in \mathscr{S}\right) \\
	&= \boldsymbol{\theta}_{Q_{0}}.
	\end{align*}
	Since $\Phi(v,1) = Q_{0}$, it follows that equation \eqref{eq:bs_valid} holds.
	
	Now, consider $\mathbf{q} = (w_{0},w_{1},\ldots,w_{d-1})\in \mathscr{V}_{\text{DB}}^{(d)}$. Then, from our construction of $\mathscr{V}_{\text{DB}}^{(d)}$, it holds that $\Phi(\mathbf{q},0) = (0,w_{0},w_{1},\ldots,w_{d-1}))=:\mathbf{q}_0^+$, and $\Phi(\mathbf{q},?) = (?,w_{0},w_{1},\ldots,w_{d-1})=:\mathbf{q}_?^+$. From equations \eqref{eq:theta_even}, \eqref{eq:Bs0} and from \eqref{eq:Bs?}, we see that $\left(B_s(\boldsymbol{\theta}_\mathbf{q},0):s\in\mathscr{S}\right) = \boldsymbol{\theta}_{\mathbf{q}^+_0}$ and $\left(B_s(\boldsymbol{\theta}_\mathbf{q},?):s\in\mathscr{S}\right) = \boldsymbol{\theta}_{\mathbf{q}^+_?}$. Thus, equation \eqref{eq:bs_valid} holds in this case too.	
\end{proof}

Now, let us define the joint distribution, $\theta(s,v):=\theta_v(s)\tilde{\pi}(v)$, where $\left(\tilde{\pi}(v):v\in \mathscr{V}^{(d)}\right)$ is the stationary distribution on the nodes of $\mathscr{G}_Q^{(d)}$ induced by $\tilde{P}$. The proof that the aperiodic input distribution, $\tilde{P}$, is BCJR-invariant, will follow from Lemma \ref{lemma:bcjr} and the following lemma:

\begin{lemma}\label{lemma:stat_dist}
	It holds that $\theta(s,v) = \tilde{\pi}(s,v)$, for all $s\in \mathscr{S}$ and $v\in \mathscr{V}^{(d)}$.
\end{lemma}
\begin{proof}
	Fix an $s^+ \in \mathscr{S}$, and a $v^{+}\in \mathscr{V}^{(d)}$. Now,
	\begin{align*}
	&\sum\limits_{s,v,x,y}\theta(s,v)P(x|s,v)P(y|x)\mathds{1}\{s^+=f(x,s)\}\mathds{1}\{v^+=\Phi(v,y)\}\\
	&= \sum\limits_{v,y}\tilde{\pi}(v)\mathds{1}\{v^+=\Phi(v,y)\}\sum\limits_{x,s}\theta_v(s)P(x|s,v)P(y|x)\times\\
	&\quad \quad \quad\quad\quad\quad\quad\quad\quad\quad\quad\quad\quad\quad\quad\quad \mathds{1}\{s^+=f(x,s)\}\\
	&\stackrel{(a)}{=}\theta_{v^+}(s^+)\sum\limits_{v,y}\tilde{\pi}(v)P(y|v)\mathds{1}\{v^+=\Phi(v,y)\}\\
	&=\theta_{v^+}(s^+)\tilde{\pi}(v^+)\\
	&= \theta(s^+,v^+),
	\end{align*}
	where (a) follows from Lemma \ref{lemma:bcjr}, since $\theta_{v^+}(s^+)P(y|q)=\sum\limits_{x,s}\theta_v(s)P(x|s,v)P(y|x)\mathds{1}\{s^+=f(x,s)\}$. Since this holds for all $(s^+,v^+) \in \mathscr{S}\times \mathscr{Y}$, it follows that $\theta(s^+,v^+) = \tilde{\pi}(s^+,v^+)$.
\end{proof}

The lemma above implies that $\boldsymbol{\theta}_v = \tilde{\pi}(\cdot|v)$, and using Lemma \ref{lemma:bcjr}, we can conclude that $\tilde{P}$ is BCJR-invariant.

\begin{remark}
	It can be shown that for a fixed value of $a$, there is a unique BCJR-invariant input distribution on $\mathscr{G}_{SQ}^{(d)}$. In particular, this means that for a fixed value of $a$, the probability\\ $P(Y=1|\mathbf{q}) = a\epsilon$, for all $\mathbf{q}\in \mathscr{V}_{\text{DB}}^{(d)}$.
\end{remark}

\section{Proof of Theorem \ref{thm:main}}
\label{sec:proof_thm}

\begin{proof}
We shall evaluate $I(X;Y|Q)$, for the $(d,\infty)$-RLL input-constrained BEC, using the BCJR-invariant input distribution $\tilde{P}$ defined in subsection \ref{subsec:BCJR_invar}, and the $Q$-graph, $\mathscr{G}_Q^{(d)}$. Now,
\begin{align*}
I(X;Y|Q) &\stackrel{(a)}{=} H(Y|Q) - h_b(\epsilon)\\
&\stackrel{(b)}{=} \bar{\epsilon}\sum_{v\in \mathscr{V}^{(d)}}\tilde{\pi}(v)h_b(P(X=1|v)) + h_b(\epsilon)- h_b(\epsilon)\\
&\stackrel{(c)}{=} \bar{\epsilon}\sum_{\mathbf{q}\in \mathscr{V}^{(d)}_{\text{DB}}}\tilde{\pi}(\mathbf{q})h_b(P(X=1|\mathbf{q}))
\end{align*}
where (a) follows from the fact that $H(Y|X,Q) = H(Y|X) = h_b(\epsilon)$, and (b) follows from the fact that $P(Y=?|v) = \epsilon$, for all $v\in \mathscr{V}^{(d)}$, with $P(Y=1|v) = \bar{\epsilon}P(X=1|v)$. We additionally use  the identity that $H(a\bar{c},\bar{a}\bar{c},c) = h_b(c)+\bar{c}h_b(a)$, for all $a,c \in [0,1]$. Finally, (c) holds since $P(X=1|Q_i) = 0$, for \\$i\in [0:d-1]$. 

Now, we note that for any $\mathbf{q}\in \mathscr{V}_{\text{DB}}^{(d)}$, $P(X=1|\mathbf{q}) = \tilde{\pi}(d|\mathbf{q})a_\mathbf{q}$, and since from Lemma \ref{lemma:stat_dist} we have that \\$\tilde{\pi}(s|\mathbf{q})= {\theta}_\mathbf{q}(s)$, for all $s\in \mathscr{S}$, it follows that $P(X=1|\mathbf{q}) = a$. Hence, we get that 

\begin{equation}\label{eq:I_eval}
I(X;Y|Q) = \bar{\epsilon}h_b(a)\sum_{\mathbf{q}\in \mathscr{V}_{\text{DB}}^{(d)}}\tilde{\pi}(\mathbf{q}).
\end{equation}
Further, from the fact that $\tilde{\pi}$ is the stationary distribution on the nodes of $\mathscr{G}_Q^{(d)}$,
\begin{align}\label{eq:stat_q}
1 &= \sum_{\mathbf{q}\in \mathscr{V}_{\text{DB}}^{(d)}}\tilde{\pi}(\mathbf{q}) + \sum_{i=0}^{d-1}\tilde{\pi}(Q_i) \notag\\
&= \sum_{\mathbf{q}\in \mathscr{V}_{\text{DB}}^{(d)}}\tilde{\pi}(\mathbf{q}) + d\tilde{\pi}(Q_0) \notag\\
&= (1+da\bar{\epsilon})\sum_{\mathbf{q}\in \mathscr{V}_{\text{DB}}^{(d)}}\tilde{\pi}(\mathbf{q}),
\end{align}
where the penultimate equality holds since $\tilde{\pi}(Q_0) = \tilde{\pi}(Q_i)$, for all $i\in [d-1]$. The last equality follows from the observation that $\tilde{\pi}(Q_0) = \sum_{\mathbf{q}\in \mathscr{V}_{\text{DB}}^{(d)}}\tilde{\pi}(\mathbf{q})P(Y=1|\mathbf{q}) = \bar{\epsilon}a\sum_{\mathbf{q}\in \mathscr{V}_{\text{DB}}^{(d)}}\tilde{\pi}(\mathbf{q})$.

Hence, from equations \eqref{eq:stat_q} and \eqref{eq:I_eval}, we obtain that $I(X;Y|Q) = \frac{h_b(a)}{da+\frac{1}{1-\epsilon}}$, and taking a maximum over all \\$a\in [0,\frac{1}{d+1}]$, we get the required result.
\end{proof}

\section{Proof of Proposition \ref{lemma:UB}}
\label{sec:proof_UB}
\begin{proof}
We shall compute the upper bound expression in theorem \ref{thm:UB} by considering the family of $Q$-graphs, $\mathcal{J}:=\{\hat{\mathscr{G}}_Q^{(d)}: d\in \mathbb{N}\}$, where $\hat{\mathscr{G}}_Q^{(d)}$ is shown in figure \ref{fig:G_hat} for some fixed $d$. Let $\hat{\mathscr{G}}_{SQ}^{(d)}$ denote the $(S,Q)$-graph corresponding to $\hat{\mathscr{G}}_Q^{(d)}$. 

\begin{figure*}[t]
	\centering
	\includegraphics[width=0.7\textwidth]{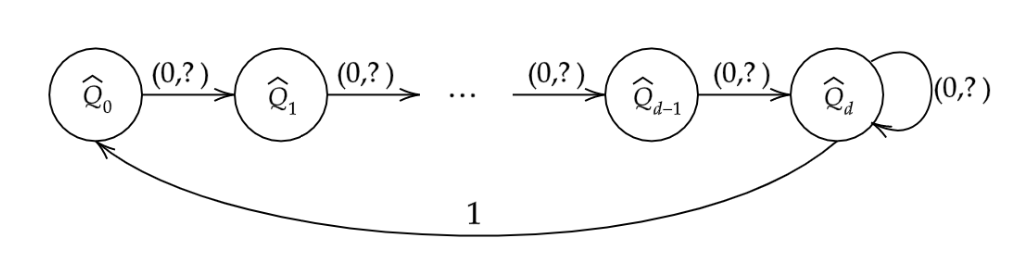}
	\caption{The $Q$-graph, $\hat{\mathscr{G}}_Q^{(d)}$. The labels on the edges denote outputs.}
	\label{fig:G_hat}
\end{figure*}

It can be shown that the single closed communicating class in $\hat{\mathscr{G}}_{SQ}^{(d)}$ is precisely the collection $\{(i,\hat{Q}_i): i\in [0:d-1]\}\cup \{(s,\hat{Q}_d):s\in [0:d]\}$ of nodes. Further, since the inputs are constrained, it holds that $P(X=1|S=s,Q=\hat{q}) = 0$, for all $(s,\hat{q}) \neq (d,\hat{Q}_d)$, with $(s,\hat{q})$ in the closed class. Therefore, the matrix of conditional input distributions, $[P(x|s,\hat{q})]$, can be parameterized by a single parameter, $p:=P(X=1|S=d,Q=\hat{Q}_d)$.

A simple computation then reveals that the stationary distribution of nodes in the closed communicating class obey:
\begin{align*}
\pi(s,\hat{Q}_s) &= \frac{p\bar{\epsilon}}{1+dp}, \ s\in [0:d-1],\\
\pi(s,\hat{Q}_d) &= \frac{p\epsilon}{1+dp}, \ s\in [0:d-1], \\
\pi(d,\hat{Q}_d) &= \frac{1}{1+dp}.
\end{align*}
Further, it holds that the conditional probability,

\begin{align*}
P(Y=1|\hat{Q}_d) &= P(Y=1,X=1,S=d|\hat{Q}_d)\\
&=p\bar{\epsilon}\frac{\pi(d,\hat{Q}_d)}{\sum_{s=0}^{d}\pi(s,\hat{Q}_d)}\\
&= \frac{p\bar{\epsilon}}{1+dp\epsilon}.
\end{align*}

Then, by Theorem \ref{thm:UB}, it holds that
\begin{align*}
C_{(d,\infty)}^{\text{fb}} &\leq \sup_{P(x|s,q)\in \Omega} I(X;Y|Q)\\
&= \sup_{P(x|s,q)\in \Omega} H(Y|Q) - h_b(\epsilon)\\
&\stackrel{(a)}{=} \max_{p\in [0,1]} \bar{\epsilon}\left(\sum_{s=0}^{d}\pi(s,\hat{Q}_d)\right)h_b\left(\frac{p}{1+dp\epsilon}\right)\\
&= \max_{p\in [0,1]} \frac{\bar{\epsilon}(1+dp\epsilon)}{1+dp}h_b\left(\frac{p}{1+dp\epsilon}\right)\\
&\stackrel{(b)}{=} \max_{\delta \in [0,\frac{1}{1+d\epsilon}]} R(\delta),
\end{align*}
where, in (a), we have used the identity that $H(a\bar{c},\bar{a}\bar{c},c) = h_b(c)+\bar{c}h_b(a)$, for all $a,c \in [0,1]$, and (b) follows by changing the variable used in the maximization to $\delta:= \frac{p}{1+dp\epsilon}$.

Now, we note that if $\epsilon^{\star}$ satisfies the equation
\[
(d\epsilon)^{\left(\frac{1}{1-\epsilon}+d\right)} = (1+d\epsilon)^d,
\]
then, the derivative evaluated at $\frac{1}{1+d\epsilon^{\star}}$,  $R^{\prime}\left(\frac{1}{1+d\epsilon^{\star}}\right)$, is equal to $0$. Further, from the proof of Corollaries \ref{coroll} in Appendix B, we observe that $R^{\prime}(\delta)$ is strictly decreasing in $\delta$, for $\delta\in [0,1]$. Hence, it follows that for $\epsilon>\epsilon^{\star}$, the derivative, $R^{\prime}\left(\frac{1}{1+d\epsilon}\right)>0$, with $R^{\prime}(0^+)>0$. 

Therefore, the non-causal capacity, which is a maximum over $[0,1]$ of $R(\delta)$, is strictly larger than the upper bound in Proposition \ref{lemma:UB}, for $\epsilon>\epsilon^{\star}$.
\end{proof}

\section{Conclusions and Future Work}
In this work, new lower bounds on the feedback capacities of the $(d,\infty)$-RLL input-constrained binary erasure channels were derived. The main idea was the construction of a family of $Q$-graphs and identifying a BCJR-invariant distribution for each graph in the family, and finally using the single-letter lower bounding methods in \cite{Single} to obtain achievable rates. The rates derived were shown to be equal to the feedback capacities, for $d=1$, and for a certain range of the channel parameter, $\epsilon$, when $d=2$. Further, numerical evaluations indicate that the lower bounds are close to the single-letter upper bounds (from \cite{Single}) derived using the same family of $Q$-graphs, for $d=2,3,4$, for all values of $\epsilon$.

Extensions of this work could look at deriving analytical expressions for upper bounds for all values of $d$, using our $Q$-graph family, and comparing them with our lower bounds. Further, we intend analyzing the structure of the optimal input distribution, derived from the dynamic programming formulation of the feedback capacity problem, which will help identify the structure of the optimal $Q$-graph.
\section*{Appendix A}
In this section, we shall prove Lemma \ref{lemma:domain}.

\begin{proof}
	Consider a node $\mathbf{q} = (w_{0},\ldots,w_{d-1})\in \mathscr{V}_{\text{DB}}^{(d)}$ with accompanying vector $\boldsymbol{\beta}$. From equation \eqref{eq:theta_even} , we obtain that
	\begin{equation*}
	\theta_\mathbf{q}(d) = 1-a\sum_{s: w_s=?}\left(\frac{1}{1-a}\right)^{\beta_s}.
	\end{equation*}
	Let us define by $n_0$, the number of times a $0$ occurs in the $d$-tuple representation of $\mathbf{q}$, i.e., $n_0:=|\{s:w_s=0\}|$. Likewise, we define $n_?:=|\{s:w_s=\ ?\}|$. Now, the following sequence of inequalities holds:
	\begin{align*}
	\sum_{s: w_s=?}\left(\frac{1}{1-a}\right)^{\beta_s} &\leq n_? \left(\frac{1}{1-a}\right)^{n_0}\\
	&\leq n_? \left(1+\frac{1}{d}\right)^{n_0},
	\end{align*}
	where the last equality holds since $a\in [0,\frac{1}{d+1}]$, and, hence, $\frac{1}{\bar{a}}\in [1,1+\frac{1}{d}]$. Now, since $n_0 = d-n_?$, it follows that
	
	\begin{align*}
	\sum_{s: w_s=?}\left(\frac{1}{1-a}\right)^{\beta_s} &\leq n_? \left(1+\frac{1}{d}\right)^{d-n_?}\\
	&= n_?\left(1+\frac{1}{d}\right)^{d}\left(\frac{d}{d+1}\right)^{n_?}\\
	&\leq d\left(1+\frac{1}{d}\right)^{d}\left(\frac{d}{d+1}\right)^d\\
	&=d,
	\end{align*}
	where the last inequality follows from the fact that the function $g(u) := u\left(\frac{d}{d+1}\right)^u$ is increasing in the interval $[0,d]$.
	
	Hence, for $a\in [0,\frac{1}{d+1}]$, we have that 
	\begin{align*}
	\theta_\mathbf{q}(d)-a &\geq 1-a-ad\\
	&=1-a(d+1)\\
	&\geq 0,
	\end{align*}
	and, therefore, $a_\mathbf{q} = \frac{a}{\theta_\mathbf{q}(d)}\in [0,1]$.
	 
\end{proof}

\section*{Appendix B}
We shall now prove the Corollaries \ref{coroll}.

\begin{proof}
	We first note that the derivative, $R^{\prime}(\cdot)$, is given by
	\begin{equation*}
	R^{\prime}(\delta) = \frac{(k+d)\log(1-\delta)-k\log\delta}{(k+d\delta)^2},
	\end{equation*} 
	where we write $\frac{1}{1-\epsilon}$ as $k$. We note that $R^\prime(\delta)$ is strictly decreasing in $\delta$.
	\begin{enumerate}
		\item It is easy to see that when $d=2$, $R^\prime\left(\frac{1}{3}\right)\leq 0$ iff $k \leq \frac{2 \log(1+1/2)}{\log 2}$, or, equivalently, iff $\epsilon \leq 1-\frac{1}{2\log(\frac{3}{2})}$. As was noted in the remark following theorem \ref{thm:main}, $R^\prime(0^+)>0$, and, hence, we have that for $\epsilon \leq 1-\frac{1}{2\log(\frac{3}{2})}$, the unique maximum of $R(\cdot)$, over $[0,1]$, occurs in the interval $[0,\frac{1}{3}]$. 
		
		In other words, for $\epsilon\leq 1-\frac{1}{2\log(\frac{3}{2})}$, our lower bound on $C_{(2,\infty)}^{\text{fb}}(\epsilon)$ coincides with $C_{(2,\infty)}^{\text{nc}}(\epsilon)$, implying that  $C_{(2,\infty)}^{\text{fb}}(\epsilon) = C_{(2,\infty)}^{\text{nc}}(\epsilon)$ for this range of $\epsilon$.
		
		Further, for $\epsilon>1-\frac{1}{2\log(\frac{3}{2})}$, we have that  $R^\prime\left(\frac{1}{3}\right)>0$, which, in turn, means that $R^\prime(\delta)>0$, for $\delta\in [0,\frac{1}{3}]$. Hence, since $R(\delta)$ is strictly increasing for $\delta\in[0,\frac{1}{3}]$, it follows that
		\begin{equation*}
		\max_{\delta\in[0,\frac{1}{d+1}]} R(\delta) = R\left(\frac{1}{3}\right).
		\end{equation*}
		\item As in the previous part, we note that  $R^\prime\left(\frac{1}{d+1}\right)>0$ iff $k>\frac{d \log(1+1/d)}{\log d}$, which, in turn, holds when $d\geq 3$, as then $\frac{d \log(1+1/d)}{\log d}<1$. Hence, for $d\geq 3$, $R^\prime(\delta)>0$, for $\delta\in [0,\frac{1}{d+1}]$, implying that $R(\cdot)$ is strictly increasing in this interval. Therefore, it follows that
		\begin{equation*}
		\max_{\delta\in[0,\frac{1}{d+1}]} R(\delta) = R\left(\frac{1}{d+1}\right).
		\end{equation*}
		Moreover, since $R^\prime(\frac{1}{d+1}) > 0$ as well, the non-causal capacity, which is the maximum of $R(\delta)$ over $\delta\in [0,1]$, is strictly greater than $R\left(\frac{1}{d+1}\right)$.
	\end{enumerate}
\end{proof}



\bibliographystyle{IEEEtran}
{\footnotesize
	\bibliography{references}}

\end{document}